\documentclass[runningheads]{llncs}
\usepackage{graphicx}
\usepackage{hyperref}
\usepackage{adjustbox}
\usepackage{multirow}

\begin{document}
\title{Cultural influence on autonomous vehicles acceptance}
\author{Chowdhury Shahriar Muzammel, Maria Spichkova, James Harland\inst{1}}
\authorrunning{C.S. Muzammel et al.}
\institute{RMIT University, Melbourne, Australia, 
\email{s3987367@student.rmit.edu.au}\\
}
\maketitle              
\begin{abstract}
Autonomous vehicles and other intelligent transport systems have been evolving rapidly and are being increasingly deployed worldwide. %
Previous work has shown that perceptions of autonomous vehicles and attitudes towards them depend on various attributes, including the respondent's age, education level and background. These findings with respect to age and educational level are generally uniform, such as showing that younger respondents are typically more accepting of autonomous vehicles, as are those with higher education levels. However the influence of factors such as culture are much less clear cut. In this paper we analyse the relationship between acceptance of autonomous vehicles and national culture by means of the well-known Hofstede cultural model.    \\
~\\
\emph{Preprint. Accepted to the 20th International Conference on Mobile and Ubiquitous Systems: Computing, Networking and Services (MobiQuitous 2023), Melbourne, Australia, November, 2023. Springer. Final version to be published by Springer (In Press).}

\end{abstract}
\section{Introduction and Motivation}
 
Autonomous vehicles (AVs) and intelligent transport systems are an increasing trend globally. These include driverless cars, trains and trams, as well as more adventurous proposals such as pilotless air taxis for up to 4 people. %
A key factor in the success of any such system, especially for mass transit, is acceptance and trust in these systems by the general public. Hence it is important to analyse and understand attitudes towards AVs, and what may be influencing them.

There are several existing studies examining this issue. These include 
a case study on public opinion on the development of an intelligent transport system in Saudi Arabia \cite{aldakkhelallah2022public,todorovic2022adoption},
surveys conducted 
in Australia \cite{aldakkhelallah2023investigation},  
Turkey \cite{meidute2021people,bacsargan2019driver}, 
Jordan \cite{abudayyeh2023perceptions}, 
Greece \cite{souris2019attitudes,gerekos2022attitudes}, etc.  
These studies demonstrate that perceptions of AVs may be influenced by a number of factors, including age, level of education, gender, and culture of the survey participants. It is noticeable that these various studies generally show a similar trend with respect to age and level of education, i.e. that younger respondents are generally more open than older ones to the usage of autonomous vehicles, and that respondents with university degrees are more accepting of AVs than those without such qualifications. However, there appears to be no such clear consensus with respect to cultural aspects. 

\emph{Contribution:} %
In this paper, we present our analysis of the acceptance of AVs on the basis of Hofstede's cultural model \cite{hofstede2010cultures}. This model has been widely used for various analyses, including those in IT and Software Engineering.\footnote{According to Google Scholar, Hofstede's work has been cited more than 243,000 times, retrieved 11/10/2023}.


\section{Background: Hofstede's model}
\label{sec:Hofstede}

There are several views on how the term \emph{culture} can be defined. It is often analysed from many perspectives, including national, regional, ethnic, religious, linguistic, and social class. One of the most popular aspects is \emph{national culture} \cite{hofstede2010cultures,trompenaars2011riding}.  
There are several approaches to define and model national  and  organisational culture \cite{house2004culture,hall1989beyond,lytle1995paradigm}.
One of the most accepted definitions and models of culture was introduced by Hofstede~\cite{hofstede2010cultures}.

Hofstede's model  covers more than 80 countries and has been applied in many culture-related studies in the field of IT and Software Engineering (and elsewhere) for comparison between cultures  and to understand the impact of cultural differences, see e.g., \cite{Borchers,spichkova2015human,spichkova2015towardsAutonom,ayed2017agile,alsanoosy2018cultural,alsanoosy2018culturalENASE,alsanoosy2019cultural,alsanoosy2019detailed,leidner2006review,spichkova2019requirements,alsanoosy2019influence,alsanoosy2020does,thanasankit2002requirements,thanasankit2002understanding,alsanoosy2020cultural,alsanoosy2020formal,alsanoosy2020identification,spichkova2021impact,alsanoosy2020exploratory,alsanoosy2020framework}.
for comparison between cultures  and to understand the impact of cultural differences~
Hofstede defines \textbf{{\textquotedblleft}national culture{\textquotedblright}} as \textit{{\textquotedblleft}the collective programming of the mind which distinguishes the members of one group or category of people from another{\textquotedblright}}. 
In this sense, national culture refers to the beliefs and values that distinguish one nation from another.

Hofstede's model consists of six dimensions: 
\begin{itemize}
\item 
\textbf{Power Distance Index (PDI):} the degree to which people within a society expect and accept that power is unequally distributed;
\item %
\textbf{Individualism/Collectivism (IDV):} the degree to which people within a society collaborate with each other;
\item %
\textbf{Masculinity/Femininity (MAS):} the degree to which the society stresses achievement or nurture;
\item 
\textbf{Uncertainty Avoidance Index (UAI):} the degree to which people within a society feel either uncomfortable or comfortable in new or unknown situations;
\item %
\textbf{Long-/Short-term Orientation (LTO):} the degree to which people within a society are linked to its own past while dealing with the present and the future challenges; and 
\item 
\textbf{Indulgence/Impulses (IND):} the degree to which people within a society have fun and enjoy life without restrictions and regulations.
\end{itemize}

Scores for each dimension vary between 0 to 100, with 50 as an average. We obtained the latest Hofstede's values corresponding to the countries associated with this study from their official website known as {\textquotedblleft}\href{https://www.hofstede-insights.com/country-comparison-tool} {Hofstede-Insights}{\textquotedblright}, as shown in the Table~\ref{tab:HofstedeDim}.
The analysis used by Hofstede is that if a score is below the average, the culture is considered low on that dimension.
For example, Australia, with a score of 38, is considered low on the PDI dimension.
The record of scores for all dimensions results in what is called the {\textquotedblleft}cultural profile{\textquotedblright} for a specific culture. Israel currently lacks a score in the IND dimension, so it is mentioned as not available (N/A). Pakistan has a score of 0 in the IND dimension, characterising it as a notably restrained society. 

\begin{table}
    \centering
    {\footnotesize{
    \begin{tabular}{|c||c|c|c|c|c|c|} \hline 
         \textbf{Country}&  \textbf{PDI}&  \textbf{IDV}&  \textbf{MAS}&  \textbf{UAI}&  \textbf{LTO}&  \textbf{IND} \\ \hline \hline
         Australia & 38 & 90 & 61 & 51 & 21 & 71 \\\hline
         Canada& 39& 80& 52& 48& 36&68\\\hline
         China& 80 & 20 & 66 & 30 & 87 & 24 \\ \hline 
         France& 68 & 71 & 43 & 86 & 63 & 48 \\ \hline
         Germany& 35 & 67 & 66 & 65 & 83 & 40 \\ \hline 
         India& 77 & 48 & 56 & 40 & 51 & 26 \\ \hline
         Indonesia& 78 & 14 & 46 & 48 & 62 & 38 \\ \hline 
         Israel& 13 & 54 & 47 & 81 & 38 & N/A \\ \hline 
         Japan& 54 & 46 & 95 & 92 & 88 & 42 \\ \hline 
         Pakistan& 55 & 14 & 50 & 70 & 50 & 0 \\ \hline
         Poland& 68 & 60 & 64 & 93 & 38 & 29 \\ \hline
         South Korea & 60 & 18 & 39 & 85 & 100 & 29\\ \hline
         UK& 35 & 89 & 66 & 35 & 51 & 69 \\ \hline 
         USA& 40 & 91 & 62 & 46 & 26 & 68 \\ \hline
    \end{tabular}
    }} 
    ~\\~
    \caption{Hofstede's PDI, IDV, MAS, UAI, LTO, IND scores}
    \label{tab:HofstedeDim}
\end{table}

\section{Cultural dimensions vs. Acceptance of AVs}
\label{sec:mapping}

Some existing studies have concluded that there is a dependency between AV acceptance and the Hofstede dimensions. %
In \cite{yun2021statistical}, the authors proposed a few hypotheses regarding the acceptance of AVs, drawing on the Hofstede cultural model.  
They conducted a case study in South Korea and compared the results with another case study~\cite{schoettle2014public} covered Australia, Japan, UK, China, USA, and India. 
The results of their cross-cultural analysis indicated that high scores of PDI and LTO \emph{positively influence} AV acceptance (i.e. higher scores of cultural dimension mean higher AV acceptance), while high scores of  IDV, MAS, UAI, and IND have a \emph{negative influence} (i.e., higher scores of cultural dimension mean lower AV acceptance). 
For simplicity of the further analysis we denote this as
\[ 
PDI^+,~ IDV^-,~ MAS^-,~ UAI^-,~ LTO^+,~ IND^-
 \]
The results of another study~\cite{taniguchi2022understanding}, which covered Japan, UK, and Germany, indicated that the population of countries with higher values of PDI and MAS dimensions will be more accepting towards AVs. The authors found no influence of IDV, UAI, LTO, or IND dimensions, i.e. 
\[ 
PDI^+,~ IDV^0,~ MAS^+,~ UAI^0,~ LTO^0,~ IND^0
 \] 
Thus, both \cite{yun2021statistical} and \cite{taniguchi2022understanding} agree regarding $PDI^+$, but disagree regarding the potential impact of other dimensions.

\begin{table}[ht!]
    \centering
    {\footnotesize{
    \begin{tabular}{|c||c|c|c|c|c|} \hline  
    \textbf{Country} &  \textbf{Ref.}&  \multicolumn{2}{c|}{\textbf{$Acc$, \%
    }
    }& \textbf{Average } & \textbf{Median}\\
     & & ~~~~ PAV ~~~~~&FAV& \textbf{$Acc$, \% } & \textbf{~deviation~~}\\\hline \hline  
        Australia& \cite{schoettle2014public} & &41&  &\\ \hline
        
        \multirow{3}{*}{Japan}&  \cite{taniguchi2022understanding}&   40.2 &30.2&\multirow{2}{*}{48.515 (PAV)}&\multirow{2}{*}{8.314 (PAV)}\\ 
        
        & \cite{khan2022risk}& 56.83&& & \\ 

        & \cite{schoettle2014public} &  &41& 35.6 (FAV)&  5.4 (FAV)\\ \hline  
        
        \multirow{2}{*}{UK}& \cite{taniguchi2022understanding}&  33.4&30.7& \multirow{2}{*}{35.35 (FAV)}& \multirow{2}{*}{4.65 (FAV)}\\ 

        & \cite{schoettle2014public} &  &40&   &   \\ \hline
        
        Germany & \cite{taniguchi2022understanding} & 29&25.8& &\\ \hline 
        
        Israel& \cite{khan2022risk}& 74.3&& &\\ \hline 

        \multirow{2}{*}{Indonesia} & \cite{sitinjak2023assessing} & & 42 (SAV)  &&\\ 
        
        & \cite{ansori2023road} & 50.3&59.2& &\\ \hline

         \multirow{2}{*}{Pakistan} & \cite{wang2021public} & & 71 (SAV)  & &\\ 
                
        & \cite{sanaullah2017autonomous} & 64&25.8& &\\ \hline

         \multirow{2}{*}{China}& \cite{wang2021public} & & 86 (SAV)  &  & \\ 

        & \cite{schoettle2014public} &  &76& & \\ \hline

        USA, Canada and Israel & \cite{haboucha2017user} & 66 &53& &\\ \hline

         \multirow{2}{*}{USA}& \cite{schoettle2014public} &  &44& \multirow{2}{*}{43 (FAV)}& \multirow{2}{*}{1 (FAV)}\\ 

        & \cite{howard2014public} &  &42& &\\ \hline
        
         India& \cite{schoettle2014public} &  &80& & \\ \hline 

        Poland & \cite{dudziak2021assessment} &  &68& &\\ \hline

        France & \cite{payre2014intention} &  &68.1& &\\ \hline

        South Korea & \cite{yun2021statistical} &  &82.2& &\\ \hline
        
    \end{tabular}
    }} 
    \caption{Acceptance of AVs: Combined results of case studies}
    \label{tab:acceptance}
\end{table}
  
We conducted a secondary study to
 analyse (wrt. Hofstede dimensions) the combined results of 12 studies covering 14 countries:  Australia, China, Canada, Germany, France, India, Indonesia, Israel, Japan, Pakistan, Poland, South Korea, UK, and USA, see \cite{schoettle2014public,yun2021statistical,taniguchi2022understanding,khan2022risk,sitinjak2023assessing,ansori2023road,wang2021public,sanaullah2017autonomous,haboucha2017user,howard2014public,dudziak2021assessment,payre2014intention}.  
 The overview of the core results are presented in Table~\ref{tab:acceptance}, where $Acc$ denotes acceptance of AVs.  
 
Please also note that two of the   studies, conducted in Indonesia \cite{sitinjak2023assessing} 
as well as in China and Pakistan \cite{wang2021public} analysed AV acceptance for very specific settings: the use of AVs for shared autonomous vehicles (SAVs), i.e. as a 
part of a shared mobility service, such as a ride-hailing service, car-sharing platform, or autonomous shuttle. 
 
In what follows, we use the following abbreviations:
\begin{itemize}
    \item 
    \emph{FAV} refers to a fully automated vehicle, which is also sometimes called a \emph{Level~5 AV}. An FAV can perform all driving tasks under all conditions without any human intervention. 
    \item 
    \emph{PAV} refers to a partially automated vehicle, which corresponds to a \emph{Level 2, 3, or 4 AV}, having varying degrees of automation but still require some level of human involvement. 
\end{itemize}
In some of the studies, the authors analyse acceptance of FAVs, however, some studies present more nuanced approached where the acceptance of partial automation is analysed as well.

Noticeably, even when the results of different case studies within the same culture/country differ, this difference is not as large as in all countries. To understand the difference, we calculated the median deviation between those common countries of our collected case studies. Among several studies covering the same culture, the USA has the least (1\%) and Japan has the greatest (8.314\%) median deviation. However, if we compare the results across all 12 studies, the differences in level of acceptance will be substantial: the lowest level has been identified in Germany (PAV: 29\%, FAV: 25.80\%) by \cite{taniguchi2022understanding}, and the highest level has been identified in  China for shared vehicles (86\%,\cite{wang2021public}) and in South Korea (82.2\%, \cite{yun2021statistical}).  
Also, if we compare results of a study \cite{haboucha2017user} conducted in USA and Israel (where the acceptance of AVs was analysed in general, over both countries) with the results of the studies conducted in USA  only \cite{schoettle2014public,howard2014public} and Israel only \cite{khan2022risk}, we see that the results of combined study are in between of the results for each separate country. 
This leads us to a conclusion matching the finding presented in \cite{yun2021statistical} that culture plays an important role in accepting AVs, as different national cultures have different level of acceptance of AVs.

\begin{figure}
\includegraphics[width=0.5\textwidth]{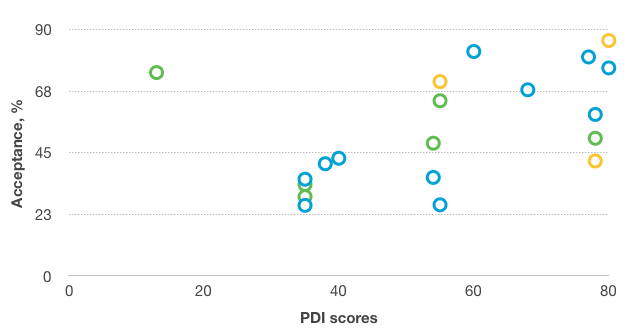}\includegraphics[width=0.5\textwidth]{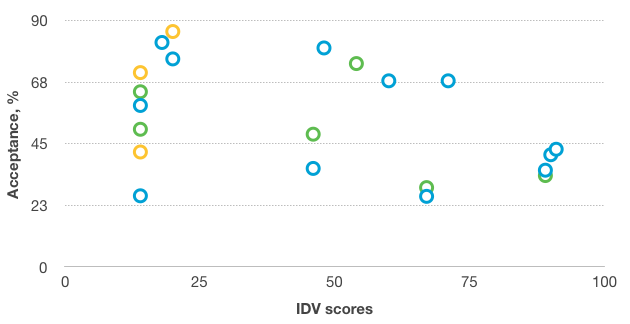}
\\

\includegraphics[width=0.5\textwidth]{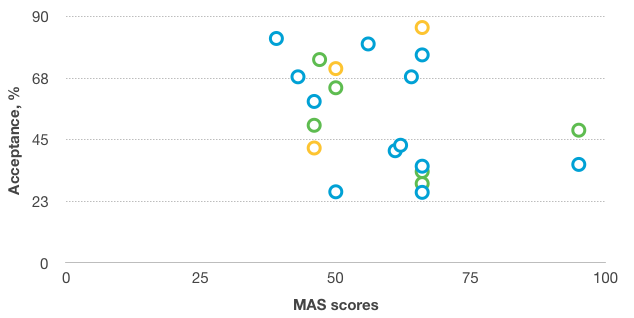}\includegraphics[width=0.5\textwidth]{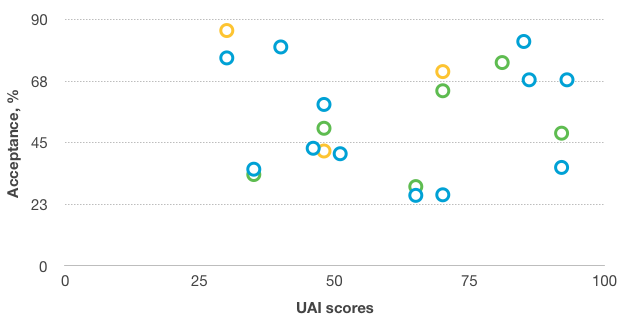}
\\

\includegraphics[width=0.5\textwidth]{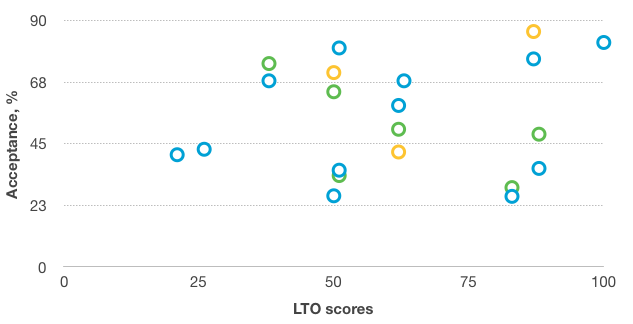}\includegraphics[width=0.5\textwidth]{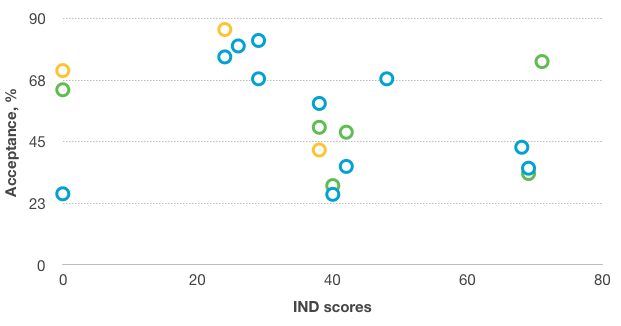}\\

\includegraphics[width=0.5\textwidth]{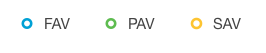}
\caption{Hofstede model scores vs. AV acceptance} 
\label{fig:correlations}
\end{figure}

 Figure~\ref{fig:correlations} presents scatter diagrams to analyse potential correlations among cultural dimensions and AV acceptance rates, where the blue, green and yellow circles present acceptance of FAV, PAV and SAV respectively, based on the data from Tables~\ref{tab:HofstedeDim} and \ref{tab:acceptance}. For the countries covered by multiple studies, we presented the average value of the acceptance rate. 
 Based on the examination of the identified case studies, we have extracted specific attributes that appear to impact the acceptance of AVs: gender-related preferences, perception of AV-related risks, social influence, and ease of use.  In what follows we analyse  potential correlations of these attributes with Hofstede's cultural dimensions. %
 
\textit{Gender-related preferences:} For Japan, the authors of  \cite{taniguchi2022understanding} identified males are more interested than females in AVs, but for the UK and Germany \cite{taniguchi2022understanding}, there is not enough evidence for gender-related preference.    
For USA \cite{saravanos2022investigating}, gender was identified as a factor influencing acceptance, contributing to 40.75\% of the variance, with males exhibiting a higher likelihood of accepting the technology than females. %
If we compare Hofstede's scores for Japan and USA with the scores for Germany (see Table~\ref{tab:HofstedeDim} for the values), there is no clear correlation between scores and the identified gender-related preferences for MAS dimension.

\textit{Risk Perception:}  
In the case study \cite{khan2022risk}, the authors compared the AV-related risk perceptions in Israel and Japan. The authors identified that in Japan, people are much more concerned about AV-related risks than in Israel. If autonomous vehicles can successfully build trust with consumers regarding their safety, consumers may perceive fewer risks associated with using autonomous vehicles.  
The authors posit that establishing trust is a crucial factor in comprehending the perceived risks linked to accepting autonomous vehicles. They believe that the level of comfort with using autonomous vehicles will increase if they can have confidence and trust in the safety of the AV technology. 
In software engineering, trust establishment is typically linked to the PDI dimensions, while risk perception is linked to UAI dimension. Therefore, a reasonable hypothesis might be to link the 
AV trust establishment  with PDI, and AV risk perception is linked with UAI. Figure~\ref{fig:correlations} also illustrates this potential, where in both cases influence is positive, i.e. 
$
 PDI^+,~ UAI^+
$.

A hypothesis from another study \cite{farzin2023autonomous} based on Iran says that risk perception has a negative impact on deciding about acceptance of AVs. In that sense, a reasonable hypothesis might be that high UAI countries should have low AV acceptance rate. According to Hofstede's scores 
and acceptance rates, 
the majority of countries having high UAIs (i.e., Australia, Indonesia, Japan, Germany, and Pakistan) has low AV acceptance rates. At the same time, France, Poland, and South Korea  have high UAI scores and exhibit high acceptance rates. Thus, there might be some correlation between AV risk perception and UAI scores. However, China has relatively low UAI (30), but according to \cite{wang2021public} its SAV= 86\% (SAV) and according to \cite{schoettle2014public} its FAV=76\%. India has medium-low UAI (40), but according to \cite{schoettle2014public}, its FAV = 80\%. 
Thus, as mentioned above, it might be a promising research direction to investigate this potential correlation.

\textit{Social influence:} Peer pressure or social influence, which represents the opinions of others, can also be another big motivation for accepting AVs~\cite{thayalan2022examining} positively or negatively. From another survey~\cite{farzin2023autonomous}, social influence plays 29\% positive role in accepting AVs for Iran. %
For USA~\cite{saravanos2022investigating}, social influence holds the second most significant influence, contributing to 44.54\% of user acceptance when it comes to autonomous vehicles (AVs). These are connected to social collaboration. People who see others drive AVs and decide to accept AVs depending on their peer's experience are relying on their peers' past experiences. 
Social influence might be potentially linked to the IDV dimension: for Collectivist societies it's typical that  opinions and votes predetermined by in-group, where Individualist Societies is typical that a personal (individual) opinion is expected. 

From Table~\ref{tab:HofstedeDim} and Figure~\ref{fig:correlations}, we can see that some countries with higher IDV (Australia, Poland, UK, USA, Germany) have lower acceptance rate, and lower IDV countries (Pakistan, Indonesia, China, South Korea, India) have higher acceptance rate.   
However, Pakistan has a very low IDV (value 0), but according to  
\cite{wang2021public} its SAV=71\%, and according to \cite{sanaullah2017autonomous} its 
PAV = 64\% and FAV=25\%. If we consider IDV values for all other countries covered by the analysed studies, the results might indicate that IDV scores might negatively influence the acceptance. 

\textit{Ease of use:}  
A hypothesis derived from a North American case study~\cite{muller2019comparing} suggests that the perceived ease of utilizing the technology positively impacts individuals' attitudes towards the adoption of AVs. 
In another case study~\cite{thayalan2022examining} conducted in Malaysia, an examination was carried out to identify the factors influencing consumer acceptance of autonomous vehicles, revealing that ease of use is one of the positive influential factors. However, this influence has no attachment to culture.

Besides analysis of the above attributes, we also would like to discuss our observations regarding the LTO dimension. 
Based on  Figure~\ref{fig:correlations}, a hypothesis on correlation between AV acceptance and LTO scores might be reasonable:  
high LTO countries might have higher acceptance rates than low LTO countries. If we analyse Hofstede's scores from Table~\ref{tab:HofstedeDim}, and acceptance rates (see Table~\ref{tab:acceptance}, we can see that majority high LTO countries (South Korea, China, France, Indonesia, Pakistan, and India) have higher acceptance rates with some exceptions (Germany and Japan). So there might be some positive influence on LTO over accepting AVs, i.e. $LTO^+$, but there is no conclusive evidence as there are few exceptions.

\section{Conclusions}
\label{sec:conclusion}

In this paper, we analysed  wrt. Hofstede cultural dimensions the combined results of 12 studies conducted in 14 countries: Australia, China, Canada, Germany, France, India, Indonesia, Israel, Japan, Pakistan, Poland, South Korea, UK, and USA.
The aim of this work was to identify \emph{potential} correlations for more focused in-depth analysis.

As result of our analysis, we identified the following potential impacts requiring further investigation:
\begin{itemize}
    \item 
    $PDI^+$, $UAI^+$, $LTO^+$: PDI/ UAI/ LTO scores might have positive impact on AV acceptance, 
    \item 
    $IDV^-$: IDV scores might have negative impact on AV acceptance,  
    \item 
    $MAS^0, IND^0$: MAS and IDV scores have no impact on AV acceptance.
\end{itemize}

We observed that even when the results of different case studies within the same culture/country differ, this difference is not as large as if we compare different countries, which highlights the correctness of the hypothesis that culture plays an important role in accepting AVs as well as potential for further work in this direction.

%
%
%
\bibliographystyle{splncs04}
%

\end{document}